\documentclass{ws-p8-50x6-00}

\begin{document}

\title{A Solution of the Generation Puzzle from Yang-Mills Duality}

\author{CHAN Hong-Mo}

\address{Rutherford Appleton Laboratory,\\
  Chilton, Didcot, Oxon, OX11 0QX, United Kingdom\\
E-mail: chanhm\,@\,v2.rl.ac.uk}  

\author{TSOU Sheung Tsun}

\address{Mathematical Institute, University of Oxford,\\
  24-29 St. Giles', Oxford OX1 3LB, United Kingdom\\E-mail: 
  tsou\,@\,maths.ox.ac.uk}

\maketitle

\abstracts{
A solution to the generation puzzle based on a nonabelian generalization 
of electric-magnetic duality is briefly reviewed.  It predicts 3 and only 
3 generations of fermions and explains the hierarchical mass spectrum as
well as the main features in both the quark and lepton mixing matrices.  
A calculation to leading perturbative order already gives reasonable values 
to about half of the Standard Model parameters.}

The complex of questions known loosely as the ``generation puzzle'' has
puzzled physcists for well over a generation.  In practical terms, it 
accounts for more than two-thirds of the twenty-odd empirical parameters 
needed to specify the Standard Model, and is thus rightly regarded by many 
as one of the most important and urgent problems facing particle physics 
today.  In this talk, we review briefly a possible solution to the puzzle 
based on a nonabelian generalization to electric-magnetic duality known 
in the literature as the DSM scheme.  For references see the list in
[I], hep-th/0007016, to which the citation numbers in this paper refer.

Let us first recall the main facts involved.  

First, the matter in our universe is made from fundamental fermions of 
4 species differing in colour or weak isospin, namely $U$-type quarks, 
$D$-type quarks, charged leptons, and neutrinos.  However, for no known 
theoretical reason, there are 3 generations to each species having the 
same quantum numbers except for their masses. 

Secondly, the masses of the 3 generations follow a markedly ``hierarchical''
pattern.  For the three charged species, the masses in MeV are 
roughly$^1$:
\begin{equation}
\left( \begin{array}{c} m_t \\ m_c \\ m_u \end{array} \right)
   = \left( \begin{array}{r} 180000 \\ 1200 \\ 4 \end{array} \right);
\left( \begin{array}{c} m_b \\ m_s \\ m_d \end{array} \right)
   = \left( \begin{array}{r} 4200 \\ 120 \\ 7 \end{array} \right);
\left( \begin{array}{c} m_\tau \\ m_\mu \\ m_e \end{array} \right)
   = \left( \begin{array}{r} 1777 \\ 105 \\ .5 \end{array} \right),
\label{UDLmasses}
\end{equation} 
where it is seen that the mass drops by one to two orders of magnitude
from generation to generation. 

Thirdly, the state vectors of the 3 generations are nearly but not exactly 
aligned between the up- and down-states.  The unitary matrix giving the
relative orientation of the down-triad to the up-triad is known as the CKM 
matrix for quarks and the MNS matrix for leptons, and present experiments
give for the absolute values of the matrix elements$^1$:
\begin{equation}
|V_{CKM}| = \left( \begin{array}{lll} 0.975\ \  & 0.220\ \  & 0.003 \\
                                    0.220\ \  & 0.974\ \  & 0.04  \\
                                    0.008\ \  & 0.04\ \   & 0.999 
\end{array} \right),
\label{CKMmat}
\end{equation}
\begin{equation}
|U_{MNS}| = \left( \begin{array}{ccc} ?\ \  & 0.4 - 0.7\ \  & 0.0 - 0.15 \\
                                    ?\ \  & ?\  & 0.45 - 0.85 \\
                                    ?\ \  & ?\ & ? \end{array} \right).
\label{MNSmat}
\end{equation}  
The matrix for quarks is close to but definitely not the identity, with 
the nonzero off-diagonal elements representing the rates of some very well 
measured hadronic proscesses.  Whereas for the leptons, the matrix is far 
from diagonal with the large off-diagonal elements representing the results 
from some recent experiments on neutrino oscillations$^{2,3}$.

By the generation puzzle, one means then not only the mystery why there
should be 3 and apparently only 3 generations of fundamental fermions,
but also why their masses and mixings should fall into such peculiar
patterns.  In current formulations of the Standard Model, these features
are taken for granted while the masses in (\ref{UDLmasses}) 
as well as the entries in (\ref{CKMmat}) and (\ref{MNSmat}) have all to 
be supplied from experiment.

Let us see now how duality helps to answer these questions.

First, the fact that there are 3 generations of fermions with very similar 
properties suggests the existence of an underlying 3-fold symmetry.  This 
new symmetry has to be broken since the 3 generations have different masses.  
The beauty of the DSM scheme is in suggesting a natural candidate for this
generation symmetry, as follows.  Electromagnetism is symmetric under the 
Hodge star operation: ${}^*\!F_{\mu\nu} = - \frac{1}{2} 
\epsilon_{\mu\nu\rho\sigma}
F^{\rho\sigma}$, which interchanges electricity and magnetism.  This implies 
in particular that the theory is invariant under a doubled gauge symmetry 
$U(1) \times \tilde{U}(1)$, where the first $U(1)$ represents the original 
(electric) gauge group while the second $\tilde{U}(1)$ represents its dual 
(magnetic) counterpart.  In terms of $U(1)$, electric charges are sources 
and magnetic charges are monopoles, but in terms of $\tilde{U}(1)$ these
charges appear instead as monopoles and sources respectively.  What
happens in nonabelian Yang-Mills fields?  It was known that under the 
Hodge star operation, Yang-Mills theory is not symmetric$^4$.
However, it was shown that if the Hodge star is replaced by a certain
generalized dual transform$^5$, then duality is 
recovered\footnote{The dual properties of Yang-Mills fields are of 
course a highly
nontrivial theoretical subject.  It has taken literally years to derive
the cited result which involved a development with Polyakov's loop space 
techniques and other things too long to be reported here.  For a brief 
outline of this work, the reader is referred to [I], and for 
more details to hep-th/9904102 and hep-th/0006178,
and original references therein.}.  
This means that the theory is again invariant under a doubled gauge
symmetry $SU(N) \times \widetilde{SU}(N)$.  In particular, for colour,
one has $SU(3) \times \widetilde{SU}(3)$.  Furthermore, it was 
shown$^{15}$ using a result of 't~Hooft$^{14}$ that given the 
confinement of colour $SU(3)$, the dual symmetry $\widetilde{SU}(3)$
is broken, so that within the colour theory itself, there is already a 
broken 3-fold symmetry ready to play the role of the generation symmetry.  
Since this symmetry already exists and needs to be physically accounted 
for in any case, it seems natural to identify it with the generation 
symmetry$^{16}$.  As a result, one concludes that there are 3 and only 3 
generations of fermions labelled by the 3 different colour magnetic 
charges, offering thus an answer to the leading question of the generation 
puzzle.

But why should the mass spectrum of fermions be hierarchical?  To answer
this, one needs to know how the $\widetilde{SU}(3)$ dual colour symmetry 
is broken which information is not given by$^{14}$.  Fortunately,
the theoretical framework developed for Yang-Mills duality$^5$
already offers candidates for the Higgs fields required in the form of
the frame vectors (complex dreibeins) $\phi_a^{(a)}; a, (a) = 1,2,3$ in 
dual colour space, which seem to play dynamical roles$^{13}$ as 
physical fields, and are $\widetilde{SU}(3)$ triplets, space-time scalars 
with finite classical lengths.  With these as Higgs fields, the following 
Yukawa coupling is suggested:
\begin{equation}
\sum_{(a)[b]} Y_{[b]} \bar{\psi}_L^a \phi_a^{(a)} \psi_R^{[b]},
\label{Yukawa}
\end{equation}
where, as in electroweak theory, we have taken left-handed fermions in
the fundamental representation, i.e. triplets, and right-handed fermions
as singlets.  In turn, the above Yukawa coupling implies a tree-level
mass matrix of the following factorized form:   
\begin{equation}
m = m_T \left( \begin{array}{c} x \\ y \\ z \end{array} \right) (x, y, z),
\label{massmat}
\end{equation}
where $(x, y, z)$ is a normalized vector with its components given by 
the vacuum expectation values of the lengths of $\phi^{(a)}, (a) = 1,2,3$.  
It follows therefore that at tree-level, $m$ has only one non-zero 
eigenvalue, offering thus the beginning of an anwer to the fermion mass 
hierarchy.

\begin{figure}
\vspace*{-4cm}
\centerline{\psfig{figure=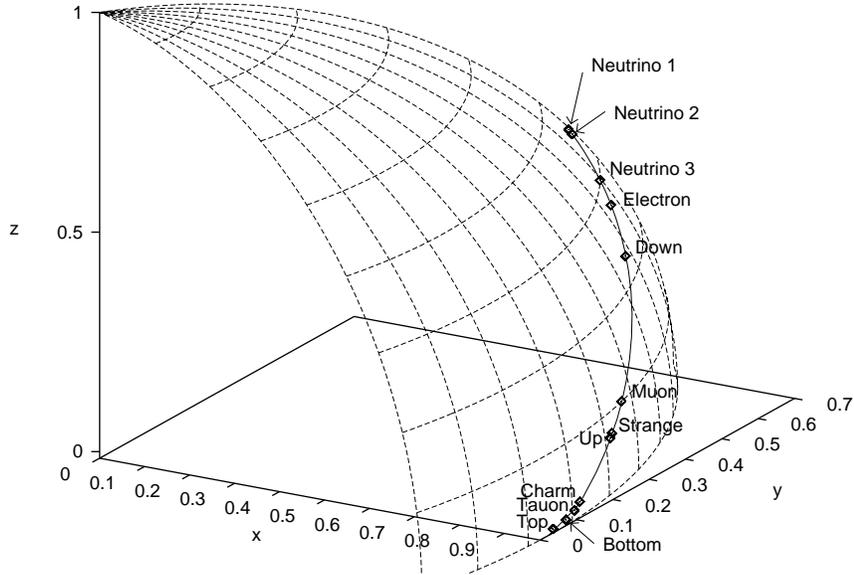,width=0.75\textwidth}}
\vspace*{-5mm}
\caption{Trajectory traced out by the vector $(x', y', z')$ in generation 
space}
\label{Jakovsphere}
\vspace*{-5mm}
\end{figure}

What happens to the mass matrix under radiative corrections?  Like many
other quantities in quantum field theory, the fermion mass matrix here
runs with changes in scale, but because of the particular way (\ref{Yukawa})
in which it is coupled, it remains factorized in form as in (\ref{massmat})
only with the vector $(x, y, z)$ now replaced by one, say $(x', y', z')$,
which varies with the scale, tracing out with changing scales a trajectory 
on the unit sphere.  The trajectory has a high energy fixed point at 
$(1,0,0)$ and a low energy fixed point at $\frac{1}{\sqrt{3}}(1,1,1)$, 
but its actual shape depends on the Higgs vev's $(x, y, z)$, and the Yukawa 
couplings, which parameters have to be fitted to data.  The result of our 
fit is shown in Figure \ref{Jakovsphere} which coincides for all fermion
species to a high accuracy$^{18,19}$.  
It is remarkable that nearly all the previously noted features of the
empirical mass spectrum and mixing matrices can now be read off from this 
figure. 

To do so, one has first to clarify some points.  The fact that the vector 
$(x', y', z')$ varies in orientation (rotates) with changing scales means 
that state vector of physical fermion states each defined at its own
mass scale, need no longer be eigenstates of the mass matrix at some other
scales.  Hence, despite the fact that the mass matrix remains factorizable 
at all scales, mass need no longer vanish for the two lower generations as 
at tree-level$^{16,20}$.  We say in this case that the mass 
``leaks'' from the highest generation into the lower ones, giving them 
small but finite masses.  Secondly, the state vectors for up- and down-states, 
being defined at different scales, need not now have the same orientation
as they did at tree-level, resulting thus in nontrivial mixing matrices.  
Notice that both lower generation masses and mixing are here governed just
by the speed at which the vector $(x', y', z')$ rotates.

With these observations made, let us turn back to examine the mass spectrum 
in (\ref{UDLmasses}).  One notices there not only that masses are dropping 
by large factors from generation to generation as expected from the leakage 
mechanism, but also that the drop factor is larger the heavier the mass, 
thus: $m_c/m_t <m_s/m_b < m_\mu/m_\tau$.  This is easily understandable, 
for the heavier the mass, the nearer the state is to the high energy fixed 
point $(1,0,0)$ as seen in Figure \ref{Jakovsphere}, and hence the slower 
the rotation and the smaller the leakage.

Next, one sees that the fact mixing is generally smaller for quarks than for 
leptons in (\ref{CKMmat}) and (\ref{MNSmat}) is also easily understood.  The 
separation between $t$ and $b$ on the trajectory in Figure \ref{Jakovsphere}  
being so much smaller than that between $\tau$ and $\nu_3$, there will be 
less rotation or disorientation between the up- and down-states for quarks 
than for leptons and hence smaller mixing also.

Finally, one sees that even the relative sizes of elements within the same 
mixing matrix, in particular, that the corner elements are exceptionally 
small in both (\ref{CKMmat}) and (\ref{MNSmat}), are readily understood
by means of some simple differential geometry$^{21}$.  To a good 
approximation, the state vectors of the three generations can be represented 
as an orthonormal (Darboux) triad at the location of the heaviest generation,
with the heaviest generation state as the radial vector to the sphere,
the second generation state as the tangent vector to the trajectory, and
the lightest generation state as the vector orthogonal to both the above.
The mixing matrix then appears just as the matrix representing the rotation 
undergone by this triad as it is transported along the trajectory from 
the location of the heaviest up-state to the heaviest down-state.  To
leading order in the distance transported, elementary differential 
geometry$^{22}$ gives this rotation matrix as:
\begin{equation}
V_{CKM} \sim \left( \begin{array}{ccc}
   1 & - \kappa_g \Delta s & - \tau_g \Delta s \\
   \kappa_g \Delta s & 1 & \kappa_n \Delta s \\
   \tau_g \Delta s & - \kappa_n \Delta s & 1  \end{array} \right),
\label{Muimat}
\end{equation}
with $\kappa_n$ being the normal curvature, $\kappa_g$ the geodesic curvature,
and $\tau_g$ the geodesic torsion of a curve on a surface.  For the unit
sphere, $\kappa_n = 1$ and $\tau_g = 0$.  From this we deduce first that the
corner elements (13 and 31) are of second order in $\Delta s$ and therefore
small compared with the others, and secondly, that the 23 and 32 elements 
are given approximately just by the transportation distance $\Delta s$, 
namely for the quark case by the distance between the top and bottom quarks 
along the trajectory, and for the lepton case by the distance between 
$\tau$ and $\nu_3$, which statement is valid to very good approximation in
(\ref{CKMmat}) and (\ref{MNSmat}) as can be easily verified in Figure 
\ref{Jakovsphere} with a piece of string. 

Of course, having constructed a detailed scheme, one can go far beyond the
qualitative discussions of the preceding paragraphs.  For example, from the 
one-loop calculation reported in$^{20}$, one obtains the numbers 
given in Table \ref{CKMtable},
\begin{table}
\caption{Predicted CKM matrix elements for both quarks and leptons}
\vspace*{-2mm}
\begin{eqnarray*}
\begin{array}{||c||c||c|c||}  
\hline \hline
Quantity & Experimental Range & Predicted & Predicted Range \\
         &                    & Central Value &             \\
\hline \hline
|V_{ud}| & 0.9745 - 0.9760 & 0.9753 & 0.9745 - 0.9762 \\ \hline
|V_{us}| & 0.217 - 0.224 & (0.2207) & input \\ \hline
|V_{ub}| & 0.0018 - 0.0045 & 0.0045 & 0.0043 - 0.0046 \\ \hline
|V_{cd}| & 0.217 - 0.224 & (0.2204) & input \\ \hline
|V_{cs}| & 0.9737 - 0.9753 & 0.9745 & 0.9733 - 0.9756 \\ \hline
|V_{cb}| & 0.036 - 0.042 & 0.0426 & 0.0354 - 0.0508 \\ \hline
|V_{td}| & 0.004 - 0.013 & 0.0138 & 0.0120 - 0.0157 \\ \hline
|V_{ts}| & 0.035 - 0.042 & 0.0406 & 0.0336 - 0.0486 \\ \hline
|V_{tb}| & 0.9991 - 0.9994 & 0.9991 & 0.9988 - 0.9994 \\ \hline
|V_{ub}/V_{cb}| & 0.08 \pm 0.02 & 0.1049 & 0.0859 - 0.1266 \\ \hline
|V_{td}/V_{ts}| & < 0.27 & 0.3391 & 0.3149 - 0.3668 \\ \hline
|V_{tb}^{*}V_{td}| & 0.0084 \pm 0.0018 & 0.0138 & 0.0120 - 0.0156 \\ \hline
   \hline
|U_{\mu3}| & 0.56 - 0.83 & 0.6658 & 0.6528 - 0.6770 \\ \hline
|U_{e3}| & 0.00 - 0.15 & 0.0678 & 0.0632 - 0.0730 \\ \hline
|U_{e2}| & 0.4 - 0.7 & 0.2266 & 0.2042 - 0.2531 \\ \hline \hline 
\end{array}
\end{eqnarray*}
\label{CKMtable}
\vspace*{-5mm}
\end{table} 
where one sees that all entries more or less overlap with the present
experimental limits, apart from one exception $U_{e2}$ which is particularly
difficult for our calculation to get correct.  And all these numbers have been 
obtained by adjusting only one parameter to the Cabibbo angle $V_{us} 
\sim V_{cd}$, the other two parameters in the scheme having already 
been fitted to fermion masses. 

One concludes thus that the DSM scheme does offer to-date a viable
solution to the generation puzzle.  Besides, with all parameters
fixed, it is now highly predictive with ramifications ranging from
rare hadron decays$^{23}$ and $\mu-e$ conversion$^{27}$, to air
showers$^{24}$ and fermion transmutation$^{29,30}$, and so far in all
these the DSM has survived existing experimental tests.

\end{document}